\documentclass[a4paper]{article}
\usepackage[psamsfonts]{amssymb}
\usepackage{amsmath}
\usepackage{epsfig}
\title{Coupled oscillators as models of quintom dark energy\vspace{1cm}}

\author{\normalsize{M. R. Setare$^{1}$\thanks{%
E-mail: rezakord@ipm.ir}  \, and \,E.~N.~Saridakis $^{2}$\thanks{%
E-mail: msaridak@phys.uoa.gr} }\\
\newline
\\
{\normalsize \it $^1$ Department of Science, Payame Noor University,
Bijar, Iran}
\\
{\normalsize \it $^2$ Department of Physics, University of Athens,
GR-15771 Athens, Greece}
\\
}

\date{}
\begin{document}
\maketitle
\newcommand{\be}{\begin{equation}}
\newcommand{\ee}{\end{equation}}
\newcommand{\bq}{\begin{eqnarray}}
\newcommand{\eq}{\end{eqnarray}}
\vspace{1cm}
\abstract

We investigate quintom cosmology in  FRW universes using
isomorphic models consisting of three coupled oscillators, one of
which carries negative kinetic energy. In particular, we examine
the cosmological paradigms of minimally-coupled massless quintom,
of two conformally-coupled massive scalars and of
conformally-coupled massive quintom, and we obtain their
qualitative characteristics as well as their quantitative
asymptotic behavior. For open or flat geometries, we find that,
independently of the specific initial conditions, the universe is
always led to an eternal expansion.

\newpage
\section{Introduction}

The type Ia supernova observations suggests that the universe is
dominated by dark energy with negative pressure, which provides
the dynamical mechanism for its accelerating expansion
\cite{{per,gar,ries}}. The strength of this acceleration is
presently a matter of debate, mainly because it depends on the
theoretical model implied when interpreting the data.

The most obvious theoretical candidate for dark energy is the
cosmological constant $\lambda$ (or vacuum energy)
\cite{Einstein:1917,cc} which has the equation of state $w=-1$.
However, as it is well known, there are two difficulties arising
from the cosmological constant scenario, namely the two famous
cosmological constant problems --- the ``fine-tuning'' and the
``cosmic coincidence'' one \cite{coincidence}. An alternative
proposal is the concept of dynamical dark energy. Such a scenario
is often realized by some scalar field mechanism and suggests that
the energy form with negative pressure is provided by a scalar
field evolving under a properly constructed potential. So far, a
large class of scalar-field dark energy models have been studied,
including quintessence \cite{quintessence}, K-essence
\cite{kessence}, tachyon \cite{tachyon}, phantom \cite{phantom},
ghost condensate \cite{ghost2}, quintom \cite{quintom}, and so
forth. It should be noted that the usual viewpoint regards
scalar-field dark energy models as an effective description of an
underlying theory of dark energy. In addition, other proposals on
dark energy include interacting dark energy models \cite{intde},
braneworld models \cite{brane}, Chaplygin gas models \cite{cg},
holographic dark energy \cite{holoext}, bulk holographic dark
energy \cite{bulkhol} and many others. In this context, scalar
fields, which may come in different forms and with a variety of
possible self-interaction potentials, constitute the dominant
(sole) form of matter in the field equations of the gravitational
theory.

 In the present work we consider scalar fields as the ones
responsible for a dark energy universe behaving as quintom, that
is obtained from an interplay of phantom and quintessence models.
 Neither of these two models alone can fulfill the transition from $w>-1$ to
$w<-1$ and vice versa. Furthermore, although in k-essence
\cite{kessence} one can have both $w\ge -1$ and $w<-1$, it has been
lately shown in \cite{Vikman1, Vikman2} that the corresponding
crossing is very unlikely to be realized during the evolution.
However, one can show \cite{{quintom},{quint2}} that considering the
combination of quintessence and phantom in a qualitatively new
model, the $-1$-transition can be fulfilled, as can be clearly seen
in \cite{quintom}. The quintom scenario of dark energy is designed
to understand the nature of dark energy with $w$ across -1. The
quintom models of dark energy differ from the quintessence, phantom
and k-essence and so on in the determination of the cosmological
evolution and the fate of the universe.

Under the assumption of a quintom field with negative kinetic
energy, the demand for a final dominance of phantom universe leads
naturally to the consideration of a minimally-coupled quintom
field, too. In the non-spatially flat universe, the problem of
quintom stability has motivated the formulation of a toy-model
\cite{carroll} consisting of three coupled oscillators, one with
negative and the others with positive-definite kinetic energy,
with the first oscillator mimicking the gravitational field while
the other two mimic the quintom (phantom and quintessence) field.
In section \ref{couploscill} we summarize the equations for the
coupled-oscillator models.  In particular in subsection
\ref{Quintomcosm} we formulate the dynamics of the massless
quintom scenario, while in \ref{Conformalcosm} we construct an
oscillator-model with positive energy which describes exactly a
conformally-coupled quintom in rescaled variables and conformal
time. In section \ref{masslessquintom} we investigate the
characteristics of the isomorphic model of the massless quintom
and in section \ref{Conformalquintom} of that corresponding to the
conformally-coupled quintom. Finally, in section \ref{Conclusions}
we summarize the obtained results and we discuss their physical
implications.

\section{Coupled oscillators as cosmological models}
\label{couploscill}

We know that toy-models, in which coupled oscillators are
utilized, are present in recent cosmological studies. In this
section we rewrite the field equations relating the quintom model
and scalar field cosmology and we show that these equations can be
reduced to the equations of a model with three coupled
oscillators. We perform our investigation in the framework of a
non-spatially-flat Friedmann-Robertson-Walker (FRW) universe which
is described by the line element \be
ds^{2}=-dt^{2}+a^{2}(t)\left(\frac{dr^2}{1-kr^2}+r^2d\Omega^{2}\right).
\label{metric} \ee

 \subsection{Quintom cosmology}
\label{Quintomcosm}

In order to describe the quintom field we use, as usual, two
scalars: $\phi$ and $\sigma$. For simplicity we assume that they
constitute the only form of matter present. The energy density and
pressure of this minimally-coupled quintom field in the metric
(\ref{metric}) are respectively:
\begin{equation}
\rho=-\frac{1}{2}\dot{\phi}^{2}+\frac{1}{2}\dot{\sigma}^{2}+V(\phi,\sigma)
\end{equation}
\begin{equation}
P=-\frac{1}{2}\dot{\phi}^{2}+\frac{1}{2}\dot{\sigma}^{2}-V(\phi,\sigma),
\end{equation}
which correspond to the normal definitions for minimally-coupled
scalars but with sign-inversion of the kinetic energy of $\phi$.
In a homogeneous and isotropic universe $\phi(t)$ and $\sigma(t)$
depend only on the comoving time $t$ and interact through the
potential $V(\phi,\sigma)$. $V(\phi,\sigma)$ should not be more
negative than a norm because, due to the final dominance of
phantom universe, the energy density $\rho$ should be non-negative
and the Hubble parameter $H$ should be real. Note that a
minimally-coupled scalar with positive kinetic energy density
$+\frac{1}{2}\dot{\phi}^{2}$ cannot exhibit $P<-\rho$ in Einstein
gravity \cite{far2}.

 For a quintom-universe described by metric
(\ref{metric}), with the $\phi$ and $\sigma$ fields as material
sources, the equations of motion are the following:
\begin{equation}
H^{2}+\frac{k}{a^{2}}=\frac{\kappa}{6}\left[\dot{\sigma}^{2}-\dot{\phi}^{2}+2V(\phi,\sigma)\right]
\label{2.1.eq1}
\end{equation}
\begin{equation}
\dot{H}+H^{2}=\frac{\kappa}{3}\left[-\dot{\sigma}^{2}+\dot{\phi}^{2}+V(\phi,\sigma)\right]
\label{2.1.eq2}
\end{equation}
\begin{equation}
\ddot{\phi}+3H\dot{\phi}-\frac{\partial V}{\partial\phi}=0
\label{2.1.eq3}
\end{equation}
\begin{equation}
\ddot{\sigma}+3H\dot{\sigma}+\frac{\partial V}{\partial\sigma}=0,
\label{2.1.eq4}
\end{equation}
where $\kappa=8\pi G$ and dots denote differentiation with respect
to the comoving time $t$. It is remarkable that only three of the
equations (\ref{2.1.eq1})-(\ref{2.1.eq4}) are independent. Indeed,
when $\dot{\phi}\neq0$ or $\dot{\sigma}\neq0$ one can derive one
of the Klein-Gordon equations (\ref{2.1.eq3}),(\ref{2.1.eq4})
using  the other and either (\ref{2.1.eq1}),(\ref{2.1.eq2}) or the
conservation equation $\dot{\rho}+3H(P+\rho)=0$ satisfied by the
quintom. Furthermore, a significant feature of equations
(\ref{2.1.eq1})-(\ref{2.1.eq4}), which determine the dynamical
evolution of the universe at hand, is that (\ref{2.1.eq3}),
(\ref{2.1.eq4}) and a combination of (\ref{2.1.eq1}) and
(\ref{2.1.eq2}), can be derived from the Lagrangian:
\begin{equation}
\mathbf{L}_0=\kappa a^{3}(\rho-P)=3a\dot{a}^{2}-3ak+\kappa
a^{3}\left[\frac{1}{2}\dot{\phi}^{2}-\frac{1}{2}\dot{\sigma}^{2}+V(\phi,\sigma)\right],\label{lagr1}
\end{equation}
or from the Hamiltonian:
\begin{equation}
\mathbf{H}_0=3a^{3}\left[H^{2}-\frac{k}{a^{2}}+\frac{\kappa}{6}\left(\dot{\phi}^{2}-\dot{\sigma}^{2}\right)-\frac{\kappa}{3}V(\phi,\sigma)\right].\label{ham1}
\end{equation}

The aforementioned simple quintom cosmological model comprises of
three degrees of freedom, one of which carries negative kinetic
energy. Therefore, we can construct a toy-model that consists of
three coupled oscillators, one with negative-definite and two with
positive-definite kinetic energy. This toy-model, although not
exact (i.e re-producing the true system), it presents the same
qualitative behavior \cite{far} and thus it can mimic the real
system described by (\ref{lagr1}) or (\ref{ham1}), in the case of
$k=0$. In the simple case of a quadratic potential
$V=m^{2}(\phi^{2}+\sigma^{2})/2$, such a model can be formulated
using the Lagrangian:
\begin{equation}
\mathbf{L}=\frac{\dot{x}^{2}}{2}-\frac{\dot{y}^{2}}{2}+\frac{\dot{z}^{2}}{2}-\frac{m_x^{2}x^{2}}{2}-
\frac{m_yy^{2}}{2}-\frac{m_zz^{2}}{2}-\frac{\mu^{2}x^{2}y^{2}}{2}-\frac{\lambda^{2}x^{2}z^{2}}{2},
\label{LL}
\end{equation}
or the associated Hamiltonian:
\begin{equation}
\mathbf{H}=\frac{\dot{x}^{2}}{2}-\frac{\dot{y}^{2}}{2}+\frac{\dot{z}^{2}}{2}+\frac{m_x^{2}x^{2}}{2}+
\frac{m_yy^{2}}{2}+\frac{m_zz^{2}}{2}+\frac{\mu^{2}x^{2}y^{2}}{2}+\frac{\lambda^{2}x^{2}z^{2}}{2}.
\end{equation}
As far as the total energy of the system remains constant, the
energy of the phantom oscillator could decrease arbitrarily, while
the energy of the other two oscillators could increase infinitely.
Hence, there is not a stable ground state for the system
\cite{{far},{carroll}}.

We can now derive the Euler-Lagrange equations from (\ref{LL}):
\begin{equation}
\ddot{x}+(m_x^{2}+\mu^{2}y^{2}+\lambda^{2}z^2)x=0
\label{quintmassive1}
\end{equation}
\begin{equation}
\ddot{y}-(m_y^{2}+\mu^{2}x^{2})y=0 \label{quintmassive2}
\end{equation}
\begin{equation}
\ddot{z}+(m_z^{2}+\lambda^{2}x^{2})z=0. \label{quintmassive3}
\end{equation}
 In the following we assume that $m_x=m_y=m_z=0$ and
$\mu^{2}=1$ and $\lambda^{2}=1$, so that equations
(\ref{quintmassive1})-(\ref{quintmassive3}) are reduced to:
\begin{equation}
\ddot{x}=-(y^{2}+z^{2})x \label{quintmassless1}
\end{equation}
\begin{equation}
\ddot{y}=x^{2}y \label{quintmassless2}
\end{equation}
\begin{equation}
\ddot{z}=-x^{2}z. \label{quintmassless3}
\end{equation}
The toy-model governed by the equations of motion
(\ref{quintmassless1})-(\ref{quintmassless3}) mimics the behavior
of that of (\ref{2.1.eq1})-(\ref{2.1.eq4}), or in other words the
massless quintom cosmological paradigm is qualitatively isomorphic
to the constructed system of coupled oscillators. This property
allows us to reveal the characteristics of the cosmological
scenario by investigating the dynamical evolution of this
oscillator system.

\subsection{Conformally-coupled scalar field cosmology}
\label{Conformalcosm}

There are many arguments supporting that in a curved space a
scalar field couples non-minimally to the Ricci curvature. The
explicit non-minimal coupling to the curvature introduces extra
terms in the equations for the scalar fields and the scale factor,
allowing  for an accelerating expansion for the universe or even
for a super-accelerating ($\dot{H}>0$) universe. Apart form the
possibility of explaining the observed recent universe
acceleration, there are additional reasons to consider such a
model. Non-minimal coupling is introduced by quantum corrections
to the action of a scalar field \cite{birell}, it is necessary for
the renormalization of the scalar field theory \cite{coleman}, and
it is even required at the classical level to preserve the
Einstein equivalence principle or to avoid causal pathologies
\cite{sonego}. Thus, in this work we consider conformally-coupled
massive scalar fields.

We use an action including scalar fields with positive,
non-minimally-coupled kinetic energy with the Ricci curvature
\cite{far}:
\begin{equation}
S=\int
d^{4}x\sqrt{-g}\left[\frac{R}{2\kappa}-\frac{\xi}{2}\left(\phi^{2}+\sigma^{2}\right)R-\frac{1}{2}g^{ab}\nabla_{a}\phi\nabla_{b}\phi
-\frac{1}{2}g^{ab}\nabla_{a}\sigma\nabla_{b}\sigma-V(\phi,\sigma)\right],
\end{equation}
where $\xi$ is a dimensionless coupling constant. We are
interested in studying the specific case where $\xi=1/6$, since
such a choice is an infrared fixed point of the relevant
renormalization group \cite{buch}. Assuming a potential form
$V=m^{2}(\phi^{2}+\sigma^{2})/2$, we derive the equations of
motion \cite{gun,fos,mor,fuk,lak,abr}:
\begin{equation}
\dot{H}+2H^{2}+\frac{k}{a^{2}}-\frac{\kappa
m^{2}}{6}\left(\phi^{2}+\sigma^{2}\right)=0 \label{2.2.eq1}
\end{equation}
\begin{equation}
\frac{\kappa}{2}(\dot{\phi}^{2}-\dot{\sigma}^{2})+\kappa
H(\phi\dot{\phi}-\sigma\dot{\sigma})-3H^{2}\left[1-\frac{\kappa}{6}\left(\phi^{2}+\sigma^{2}\right)\right]
+\frac{\kappa m^{2}}{2}\left(\phi^{2}+\sigma^{2}\right)=0
 \label{2.2.eq2}
\end{equation}
\begin{equation}
\ddot{\phi}+3H\dot{\phi}+\frac{R}{6}\phi+m^{2}\phi=0
 \label{2.2.eq3}
\end{equation}
\begin{equation}
\ddot{\sigma}+3H\dot{\sigma}-\frac{R}{6}\sigma-m^{2}\sigma=0,
 \label{2.2.eq4}
\end{equation}
along with the Hamiltonian constraint as the first fundamental
FRW equation:
\begin{equation}
H^{2}+\frac{k}{a^{2}}=\frac{\kappa}{3}\rho.
\end{equation}
The effective energy density and pressure of the scalar fields,
which guarantee energy conservation, are given by \cite{bell}:
\begin{eqnarray}
\rho=\frac{1}{2}\left(\dot{\phi}^{2}-\dot{\sigma}^{2}\right)+\frac{m^{2}}{2}\left(\phi^{2}+\sigma^{2}\right)+\frac{k}{2a^2}\left(\phi^{2}+\sigma^{2}\right)+
\frac{1}{2}H\phi(H\phi+2\dot{\phi})
+\frac{1}{2}H\sigma(H\sigma-2\dot{\sigma}) \label{rhocompl}
\end{eqnarray}
and
\begin{eqnarray}
P=
\frac{1}{2}\left(\dot{\phi}^{2}-\dot{\sigma}^{2}\right)-\frac{m^{2}}{2}\left(\phi^{2}+\sigma^{2}\right)-\frac{k}{6a^2}\left(\phi^{2}+\sigma^{2}\right)
-\frac{1}{6}\left[4H(\phi\dot{\phi}-\sigma\dot{\sigma})+\right.\nonumber\\
\left. +2(\dot{\phi}^{2}-\dot{\sigma}^{2})
+2\phi\ddot{\phi}-2\sigma\ddot{\sigma}+(2\dot{H}+3H^{2})(\phi^{2}+\sigma^{2})\right].
\end{eqnarray}

Equations (\ref{2.2.eq1})-(\ref{2.2.eq4}) and expression
(\ref{rhocompl}) for $\rho$ are evidently complicated. However, we
can reduce the problem to a system of three coupled oscillators
with sharply defined energies in a fictitious Minkowski space,
repeating the same steps as in the previous subsection. Indeed,
inserting the auxiliary variables:
\begin{equation} x\equiv ma,\
y\equiv\sqrt{\frac{\kappa m^{2}}{6}}\,a\phi, \
z\equiv\sqrt{\frac{\kappa m^{2}}{6}}\,a\sigma,
 \label{varchange}
\end{equation}
introducing the rescaled (conformal) time $\eta$ given as
$dt=ad\eta$, and assuming a quadratic potential
$V=m^{2}(\phi^{2}+\sigma^{2})/2$, the equations of motion are
transformed to \cite{cal,roch,cast}:
\begin{equation}
x''=(y^{2}+z^{2}-k)x \label{confscalar1}
\end{equation}
\begin{equation}
y''=-x^{2}y \label{confscalar2}
\end{equation}
\begin{equation}
z''=x^{2}z, \label{confscalar3}
\end{equation}
where primes denote differentiation with respect to $\eta$. Note
that these equations can be obtained from the Lagrangian:
\begin{equation}
\mathbf{L}_1=-\frac{1}{2}(x')^{2}+\frac{1}{2}(y')^{2}-\frac{1}{2}(z')^{2}-\frac{1}{2}x^{2}y^{2}-\frac{1}{2}x^{2}z^{2}+\frac{1}{2}kx^{2},
\label{L1}
\end{equation}
or from the Hamiltonian:
\begin{equation}
\mathbf{H}_1=-\frac{1}{2}(x')^{2}+\frac{1}{2}(y')^{2}-\frac{1}{2}(z')^{2}+\frac{1}{2}x^{2}y^{2}+\frac{1}{2}x^{2}z^{2}-\frac{1}{2}kx^{2}.
\end{equation}
Clearly, the system of equations
(\ref{confscalar1})-(\ref{confscalar3}) is isomorphic to that of
(\ref{2.2.eq1})-(\ref{2.2.eq4}). We mention here that $y$ and $z$
are a mixture of the gravitational and scalar-field degrees of
freedom, whereas $x$ is associated solely with gravity.
Furthermore, as it is required by the first relation of
(\ref{varchange}), the restriction $x>0$ must be applied.

\section{Dynamical behavior of massless-quintom oscillator-model}
\label{masslessquintom}

In the previous section we formulated the use of oscillator-models
in investigating the dynamical behavior of cosmological paradigms.
For a spatially flat universe under the massless phantom scenario,
the associated system of two coupled oscillators was studied by
Castagnino {\it{et al}} \cite{cast} (see also \cite{far}). In this
section we examine the evolution characteristics of the
oscillator-model corresponding to the massless quintom, which
consists of three degrees of freedom and was formulated in
subsection \ref{Quintomcosm}.  In this case the phase-space
stability analysis leads to the following results:

-Firstly, we extract  the fixed points of the system
(\ref{quintmassless1})-(\ref{quintmassless3}), defined as those
points where the velocities of the oscillators are zero. It is
easy to see that these are simply the loci $(x_0,0,0)$ and
$(0,y_0,z_0)$. In order to examine the stability of these fixed
points, as usual we calculate the partial derivatives of the right
hand sides of the system
(\ref{quintmassless1})-(\ref{quintmassless3}) at these points, and
we extract the eigenvalues of the corresponding matrix
\cite{Hilbert}. Since at least one of the eigenvalues is always
positive, we conclude that the fixed points are all unstable. In
fact, a three-oscillator system with one of them having
negative-definite kinetic energy, does not possess positions of
equilibrium.

-Due to these instabilities, apart from the fixed points, all the
orbits in the phase-space go to infinity as $t\rightarrow\infty$.
In particular, $x(t)\rightarrow\infty$ monotonically while $y(t)$
and $z(t)\rightarrow 0$ oscillating, a behavior which is
independent of the choice of initial conditions. This can be
easily verified by simple numerical investigations.

-Cycles (periodic orbits) are not possible. However, chaotic
dynamics may appear. This is a robust result for the case of the
quadratic (or equivalently the Yang-Mills) potential, and arises
from the corresponding extensive studies of the literature (see
for example \cite{Chaos}).

-In general, as it has been shown in \cite{far,cast}, and taking
into account the invariance under the transformation
$\left(x,y,z\right) \rightarrow \left(-x,y,z\right) $ and $ \left(
x, y,z \right) \rightarrow  \left( x,- y,z \right) $ and
$\left(x,y,z\right) \rightarrow \left(x,y,-z\right) $, we can
consider an asymptotic solution as:
\begin{equation}
y(t)\approx
z(t)\approx\frac{\sqrt{2}}{t}\sin\left(\frac{t^{3}}{3}\right)
\label{asym.massless1}
\end{equation}
\begin{equation}
x(t)\approx t^{2}.\label{asym.massless2}
\end{equation}
In this case the kinetic energy of the $y$ and $z$-oscillators
(for large times) is
\begin{equation}
K^{(y,z)}=\frac{(\dot{y}^{2})}{2}\approx
t^{2}\cos^{2}\left(\frac{t^{3}}{3}\right).
\end{equation}
We can see that this expression oscillates with divergent
amplitude, while the kinetic energy of the $x$-oscillator,
$K^{(x)}=-\frac{(\dot{x}^{2})}{2}\approx-2t^{2}\rightarrow-\infty$.
This behavior corresponds to the instability described in
\cite{carroll}.

Having examined the system characteristics using the auxiliary
degrees of freedom (\ref{varchange}) we can now transform back to
the physical variables $a$, $\phi$ and $\sigma$, under the
restriction $x>0$, noting the necessary inversion between
variables indicated in \cite{far}. Doing so our results can be
re-written as follows:

-The fixed points of the cosmological model are just $(a_0,0,0)$
(we discard the family $(0,y_0,z_0)$ since it corresponds to the
non-physical case of a universe with zero scale factor). Thus,
they correspond to Minkowski spaces with constant scale factor,
i.e without expansion. The fact that they are unstable, i.e the
absence of attractors, implies that an arbitrary small
perturbation can lead to the aforementioned diverging orbits.

-In particular, the previously analyzed behavior of $x(t)$, $y(t)$
and $z(t)$ implies that the scale factor $a$ goes to infinity,
while the scalar fields are oscillating, in general with a varying
period. Thus, we conclude that we always acquire an expanding
universe. This is true even in the case where the system lies
initially in one of the fixed points (constant scale factor and
zero derivative), since an arbitrary small perturbation is
sufficient to lead it to the aforementioned expanding case.
Finally, as was shown in \cite{far}, the asymptotic solutions
(\ref{asym.massless1}),(\ref{asym.massless2}) correspond to a
matter-dominated universe with $a(t)=a_0t^{2/3}$. That is, in this
model there is no mechanism that can end the expansion, either by
reversing it to contraction or by stabilizing the universe to a
steady-state type. Thus, increasing dilution and the ``thermal
death'' of the universe are inevitable.

-Periodic orbits do not exist. That is, a massless quintom cannot
drive cyclic universes \cite{cyclic}. However, since chaotic
behavior is possible, we conclude that we can obtain chaotic
cosmological evolution. Indeed, FRW cosmologies are known to
present chaotic behavior \cite{Chaos.cosmol}.

\section{Dynamical behavior of conformally-coupled quintom oscillator-model}
\label{Conformalquintom}

Let us now examine the dynamical characteristics of a massive
conformally-coupled quintom, extending the associated
oscillator-model formulated in subsection \ref{Conformalcosm} (see
\cite{szy} for the corresponding problem for a phantom field). The
Klein-Gordon equations of the scalar fields are:
\begin{equation}
\ddot{\phi}+3H\dot{\phi}-m^{2}\phi-\xi R\phi=0
\end{equation}
\begin{equation}
\ddot{\sigma}+3H\dot{\sigma}+m^{2}\sigma+\xi R\sigma=0.
\end{equation}
In terms of the auxiliary variables $x$, $y$ and $z$ defined in
(\ref{varchange}) and the conformal time $dt=ad\eta$, these field
equations can be derived from the Lagrangian:
\begin{equation}
\mathbf{L}_2=\frac{x'^{2}}{2}+\frac{y'^{2}}{2}+\frac{z'^{2}}{2}+\frac{x^{2}y^{2}}{2}+\frac{x^{2}z^{2}}{2}-\frac{1}{2}kx^{2},
\label{L2}
\end{equation}
or from the Hamiltonian:
\begin{equation}
\mathbf{H}_2=\frac{x'^{2}}{2}+\frac{y'^{2}}{2}+\frac{z'^{2}}{2}-\frac{x^{2}y^{2}}{2}-\frac{x^{2}z^{2}}{2}+\frac{1}{2}kx^{2}.
\end{equation}
Note that the Lagrangian (\ref{L2}) is equivalent to
$\mathbf{L}_1$ of equation (\ref{L1}) (apart from an overall sign)
provided that the conformally-coupled scalar is turned into a
phantom field. The dynamical system in this case is:
\begin{equation}
x''=(y^{2}+z^{2}-k)x \label{confquint1}
\end{equation}
\begin{equation}
y''=x^{2}y \label{confquint2}
\end{equation}
\begin{equation}
z''=x^{2}z. \label{confquint3}
\end{equation}
 In order to perform a stability analysis, we have to distinguish between the various $k$-cases.

-Firstly, in the case of a flat universe ($k=0$) we can see that
the fixed points are just $(x_0,0,0)$, while for $k=1$ they are
 $(x_0,\pm\nu,\pm\sqrt{1-\nu^2})$. Both these loci
correspond to empty Minkowski spaces. However, for the case of a
closed universe ($k=-1$) there are not fixed points, apart from
the trivial case of a zero-scale-factor universe.

-It is straightforward to see that these fixed points are
unstable, and that all the (non-stationary) orbits in the
phase-space go to infinity as $t\rightarrow\infty$ \cite{cast}. In
particular, numerical investigation shows that for $k=0,1$,
$x(t)\rightarrow\infty$ monotonically while $y(t)$ and
$z(t)\rightarrow 0$ oscillating. In fact, the oscillatory nature
of the solutions for $k=0$ is a general feature of non-minimally
coupled scalar fields with $\xi>0$, even in the case where
$\xi\neq1/6$ \cite{mor}. Finally, in the case $k=-1$ there is not
a qualitatively general asymptotic behavior and the system
evolution can be more complicated. In this case, as is confirmed
by numerical integration, the scale factor can sometimes decrease.

-Thus, in the cases of open and flat universes, all the solutions
that are not stationary represent universes expanding to infinity.
However, due to the instabilities, we conclude that even an
arbitrary small perturbation can bring the universe out of
stationarity, and lead it to an everlasting expansion. Therefore,
independently of the initial conditions, we always obtain an
expanding universe. The expansion cannot be reversed or end, and a
complete dilution is inevitable. Finally, note that contrary to
the case of the previous section, there is not a specific
quantitatively asymptotic behavior. Thus, the Hubble parameter can
be constant, or increasing, leading to an accelerating or even
super-accelerating universe. On the other hand, for the case of a
closed universe, the system evolution is more complex and
difficult to be outlined.

-Closed orbits in phase-space do not exist, that is a massive
conformally-coupled quintom cannot bring about a cyclic universe
\cite{cyclic}. However, due to the quadratic nature of the
potential, chaotic behavior is possible, which, as we have already
mentioned, is expected in FRW cosmologies \cite{Chaos.cosmol}.
Finally, chaoticity is more easily obtained in the case of a
closed universe ($k=-1$), where the dynamics of the system can be
more complicated.

\section{Conclusions}
\label{Conclusions}

In this work we investigate the evolution characteristics of
quintom universes, using oscillator-models as an isomorphic
description. Indeed, one can construct such models which present
all the relevant information, examine their dynamical behavior
under stability analysis, and finally transform the results back
to the cosmological picture. In particular, in \cite{cast} the
authors have constructed a toy-model with two degrees of freedom
in order to isomorphically describe the cosmological paradigm of a
minimally-coupled massless phantom field and a massless graviton
in physical time. In such a case the perturbations of the model
are unstable and this feature led the authors of \cite{carroll} to
extend it by the insertion of a negative mass term to the
potential, necessary for the stabilization of perturbations (the
model of Castagnino {\it{et al.}} of \cite{cast} could be
recovered by forcing the phantom to be massless).

The dynamical system (\ref{quintmassless1})-(\ref{quintmassless3})
of the present work constitutes a toy-model for a minimally
coupled massless quintom field and a massless graviton in physical
time, since in this case the  isomorphic description requires
three degrees of freedom. Expressing the results in the physical
framework we find that the fixed points of the system are just the
Minkowski spaces, but are unstable under, even arbitrarily small,
perturbations. Moreover, there are no attractor points and all the
phase-space orbits go to infinity with increasing time. In
particular, in this asymptotic case we obtain a matter-dominated
universe with $a(t)=a_0t^{2/3}$. Thus, we conclude that,
independently of the initial conditions, the universe is always
led to an eternal expansion.

The dynamical system (\ref{confscalar1})-(\ref{confscalar3})
describes two conformally-coupled massive scalar fields. For flat
and open geometries, all the fixed points are empty Minkowski
spaces and they are unstable. With increasing time the scalar
fields go to zero in an oscillatory way.  The obtained evolution
reveals that the matter content dilutes progressively up to its
complete evanescence while the universe expands. Such a behavior
is consistent with the phenomenology of non-minimally-coupled
scalar fields. In addition, periodic orbits are not present, and
thus this cosmological paradigm cannot drive cyclic universes.
However, chaotic dynamics may arise, a feature already known for
FRW cosmologies. On the other hand, for a closed universe, the
dynamics of the system is more complex, and it is hard to
determine even a general qualitative behavior.

The dynamical system (\ref{confquint1})-(\ref{confquint3})
corresponds to a conformally-coupled massive quintom field.
Investigating the behavior of its phase-space we see that for a
flat or open universe the fixed points represent Minkowski spaces,
which are unstable under perturbations. Thus the universe is
expanding to infinity. However, this scenario allows for a general
(unspecified) Hubble parameter, i.e for either accelerating or
even super-accelerating universe. Closed orbits, corresponding to
cyclic behavior, do not appear but chaoticity does. Finally, for a
closed universe the system evolution can be more complicated.

In conclusion, we observe that we can extract qualitative as well
as (asymptotically) quantitative characteristics of various
quintom paradigms, by investigating the corresponding isomorphic
coupled-oscillator models. The stability analysis and the obtained
asymptotic behaviors show that the quintom scenario is consistent
with cosmological observations.


\begin{thebibliography}{99}

 \bibitem{per} S. Perlmutter {\it{et al.}}, Astrophys. J. {\bf 517}, 565,
(1999).

\bibitem{gar} P. M. Garnavich {\it{et al.}}, Astrophys. J, {\bf 493}, L53,
(1998).

\bibitem{ries} A. G. Riess {\it{et al.}}, Astron. J. {\bf 116}, 1009, (1998).

\bibitem{Einstein:1917} A. Einstein, Sitzungsber. K. Preuss.
Akad. Wiss. 142 (1917) [{\it The Principle of Relativity} (Dover,
New York, 1952), p. 177].

\bibitem{cc}S.~Weinberg,
Rev.\ Mod.\ Phys.\  {\bf 61}, 1 (1989);\\
V.~Sahni and A.~A.~Starobinsky, Int.\ J.\ Mod.\ Phys.\ D {\bf 9},
373 (2000) [astro-ph/9904398];\\
  S.~M.~Carroll,
Living\ Rev.\ Rel.\ {\bf 4}, 1 (2001) [astro-ph/0004075];\\
P.~J.~E.~Peebles and B.~Ratra,
Rev.\ Mod.\ Phys.\  {\bf 75}, 559 (2003) [astro-ph/0207347];\\
T.~Padmanabhan, Phys.\ Rept.\  {\bf 380}, 235 (2003)
[hep-th/0212290].

\bibitem{coincidence} P.~J.~Steinhardt, in {\it Critical Problems in
Physics}, edited by V.~L.~Fitch and D.~R.~Marlow (Princeton
University Press, Princeton, NJ, 1997).

\bibitem{quintessence}
P.~J.~E.~Peebles and B.~Ratra,
Astrophys.\ J.\  {\bf 325}, L17 (1988);\\
B.~Ratra and P.~J.~E.~Peebles,
Phys.\ Rev.\ D {\bf 37}, 3406 (1988);\\
C.~Wetterich,
Nucl.\ Phys.\ B {\bf 302}, 668 (1988);\\
R.~R.~Caldwell, R.~Dave and P.~J.~Steinhardt, Phys.\ Rev.\ Lett.\
{\bf 80}, 1582 (1998)
[astro-ph/9708069];\\
A.~R.~Liddle and R.~J.~Scherrer, Phys.\ Rev.\ D {\bf 59}, 023509
(1999)
[astro-ph/9809272];\\
I.~Zlatev, L.~M.~Wang and P.~J.~Steinhardt, Phys.\ Rev.\ Lett.\
{\bf 82}, 896 (1999)
[astro-ph/9807002];\\
Z. G. Huang,  H. Q. Lu, and W. Fang, Class. Quant. Grav. {\bf23},
6215, (2006), [hep-th/0604160].

\bibitem{kessence}
C.~Armendariz-Picon, V.~F.~Mukhanov and P.~J.~Steinhardt, Phys.\
Rev.\ Lett.\  {\bf 85}, 4438 (2000)
[astro-ph/0004134];\\
C.~Armendariz-Picon, V.~F.~Mukhanov and P.~J.~Steinhardt, Phys.\
Rev.\ D {\bf 63}, 103510 (2001) [astro-ph/0006373].

\bibitem{tachyon}
  A.~Sen,
  JHEP {\bf 0207}, 065 (2002)
  [hep-th/0203265];\\
    T.~Padmanabhan,
    Phys.\ Rev.\ D {\bf 66}, 021301 (2002)
  [hep-th/0204150];\\
M. R. Setare, Phys. Lett. B {\bf 653}, 116, (2007).

 \bibitem{phantom}
  R.~R.~Caldwell,
    Phys.\ Lett.\ B {\bf 545}, 23 (2002)
  [astro-ph/9908168];\\
    R.~R.~Caldwell, M.~Kamionkowski and N.~N.~Weinberg,
    Phys.\ Rev.\ Lett.\  {\bf 91}, 071301 (2003)
  [astro-ph/0302506];\\
   S. Nojiri and S. D.
Odintsov, Phys. Lett. B {\bf562}, 147 (2003) [hep-th/0303117];\\
S. Nojiri and S. D. Odintsov, Phys. Lett. B {\bf565}, 1 (2003) [hep-th/0304131];\\
M. R. Setare,  Eur. Phys. J.  C {\bf 50}, 991, (2007).

\bibitem{ghost2}
  F.~Piazza and S.~Tsujikawa,
    JCAP {\bf 0407}, 004 (2004)
  [hep-th/0405054].

\bibitem{quintom}
  B.~Feng, X.~L.~Wang and X.~M.~Zhang,
    Phys.\ Lett.\ B {\bf 607}, 35 (2005)
  [astro-ph/0404224];\\
    A. Anisimov, E. Babichev and A. Vikman,
   JCAP {\bf 0506}, 006 (2005)
  [astro-ph/0504560];\\
M. R. Setare, Phys. Lett. B {\bf641}, 130 (2006) [hep-th/0611165];\\
E. Elizalde , S. Nojiri, and S. D. Odintsov, Phys. Rev. {\bf D70},
043539 (2004) [hep-th/0405034];\\
S. Nojiri, S. D. Odintsov, and S. Tsujikawa, Phys. Rev. {\bf D71},
063004 (2005) [hep-th/0501025].

\bibitem{intde}
  L.~Amendola,
    Phys.\ Rev.\ D {\bf 62}, 043511 (2000)
  [astro-ph/9908023];\\
   D.~Comelli, M.~Pietroni and A.~Riotto,
    Phys.\ Lett.\ B {\bf 571}, 115 (2003)
  [hep-ph/0302080];\\
M. R. Setare,  JCAP {\bf0701}, 023 (2007) [hep-th/0701242].

\bibitem{brane}
  C.~Deffayet, G.~R.~Dvali and G.~Gabadadze,
   Phys.\ Rev.\ D {\bf 65}, 044023 (2002)
  [astro-ph/0105068];\\
    V.~Sahni and Y.~Shtanov,
   JCAP {\bf 0311}, 014 (2003)
  [astro-ph/0202346];\\
  M. R. Setare,  Phys. Lett. {\bf B642}, 421, (2006);\\
  F.~K.~Diakonos and E.~N.~Saridakis
  [arXiv:hep-th/0708.3143].

  \bibitem{cg}
  A.~Y.~Kamenshchik, U.~Moschella and V.~Pasquier,
    Phys.\ Lett.\ B {\bf 511}, 265 (2001)
  [gr-qc/0103004];\\
M. R. Setare, Phys. Lett.  B {\bf 648}, 329, (2007);\\ M. R.
Setare, Phys. Lett. B {\bf 654}, 1, (2007).

\bibitem{holoext}
M.~Li, Phys.\ Lett.\ B {\bf 603}, 1 (2004);\\
K.~Enqvist and M.~S.~Sloth, Phys.\ Rev.\ Lett.\  {\bf 93}, 221302
(2004) [hep-th/0406019];\\
D.~Pavon and W.~Zimdahl, Phys.\ Lett.\ B {\bf 628}, 206 (2005)
[gr-qc/0505020];\\
E.~Elizalde, S.~Nojiri, S.~D.~Odintsov and P.~Wang, Phys.\ Rev.\ D
{\bf 71}, 103504 (2005) [hep-th/0502082];\\
B.~Hu and Y.~Ling, Phys.\ Rev.\ D {\bf 73}, 123510 (2006)
[hep-th/0601093];\\
M. R. Setare, Phys. Lett. B {\bf 644}, 99, (2007) [hep-th/0610190];\\
M. R. Setare, J. Zhang, X. Zhang, JCAP {\bf0703}, 007 (2007)
[gr-qc/0611084].

\bibitem{bulkhol}
E.~N.~Saridakis, Phys. Lett. B {\bf 660}, 138 (2008)
[arXiv:hep-th/0712.2228];\\
E.~N.~Saridakis, JCAP {\bf{0804}}, 020, (2008)
[arXiv:astro-ph/0712.2672];\\
E.~N.~Saridakis, Phys. Lett. B {\bf 661}, 335 (2008)
 [arXiv:gr-qc/0712.3806].
\bibitem{Vikman1}
A. Vikman, Phys.\ Rev.\ D {\bf 71}, 023515 (2005).

\bibitem{Vikman2} G-B.~Zhao,
J-Q.~Xia , M.~Li, B.~Feng, X.~Zhang , Phys. Rev. D {\bf 72}, 123515
(2005).
\bibitem{quint2}M. Li, B. Feng, and X. Zhang, JCAP {\bf 0512}, 002 (2005); J-Q. Xia, G-B. Zhao, B. Feng and X. Zhang,
JCAP 0609, 015 (2006); B. Feng, M. Li , Y-S. Piao and X. Zhang,
Phys. Lett. B 634, 101 (2006);
 M. R. Setare, J. Sadeghi, and A. R. Amani, Phys. Lett. B 660, 299
(2008); J. Sadeghi , M. R. Setare, A. Banijamali, and F. Milani,
Phys. Lett. B{\bf 662}, 92, (2008).
\bibitem{carroll}
S.~M.~Carroll, M.~Hoffman, and M.~Trodden, Phys.\ Rev.\ D {\bf 68},
023509 (2003).

\bibitem{far2}
V.~Faraoni, Int.\ J.\ Mod.\ Phys.\ D {\bf 11}, 471 (2002).

\bibitem{far}
V.~Faraoni, Phys. Rev. D {\bf 69}, 123520 (2004) [gr-qc/0404078].

\bibitem{birell}
N.~D.~Birrell and  P.~C.~W.~Davies,  Phys. Rev. D {\bf 22}, 322
(1980);\\ N.~D.~Birrell and P.~C.~W.~Davies, {\em Quantum Fields
in Curved Space} (Cambridge University Press, Cambridge, England,
1980);\\ B.~L.~Nelson and P.~Panangaden,  Phys. Rev. D {\bf 25},
1019 (1982).

\bibitem{coleman}
 C.~G.~Callan Jr., S.~Coleman and R.~Jackiw,  Ann.
Phys. (NY) {\bf 59}, 42 (1970).

\bibitem{sonego}
 S. Sonego and V. Faraoni,   Class.
Quant. Grav. {\bf 10}, 1185 (1993);\\   R.~B.~Mann and
R.~G.~McLenaghan eds. (World Scientific, Singapore, 1994).

\bibitem{buch}
I.~Buchbinder, S.~D.~Odintsov and I.~Shapiro, {\it {Effective
Action In
Quantom Gravity}} (IOP Publishing, Bristol, 1992);\\
L.~Parker and D.~J.~Toms, Phys.\ Rev.\ D {\bf 32}, 1409 (1985).

\bibitem{gun}
E.~Gunzig {\it{et al.}}, Class.\ Quant.\ Grav. {\bf 63}, 067301
(2001).

\bibitem{fos}
S.~Foster, Class.\ Quant.\ Grav.\ {\bf 15}, 3485 (1998);\\
S. Blanco {\it{et al.}}, Gen.\ Rel.\ Grav. {\bf 26}, 1131 (1994).

\bibitem{mor}
M.~Morikawa, AStrophys.\ J.\ (Lett.)\ {\bf 362} L37 (1990);
Astrophys.\ J.\ {\bf 369}, 20 (1991).

\bibitem{fuk}
T.~Fukuyama {\it{et al.}}, Int.\ J.\ Mod.\ Phys.\ D {\bf 6} 69
(1997).

\bibitem{lak}
M.~Lakshamanan and R.~ Sahadevan, Phys.\ Rep\ {\bf 224}, 1 (1993).

\bibitem{abr}
L.~R.~ Abramo {\it{et al.}}, Phys.\ Rev.\ D\ {\bf 67}, 027301
(2003).

\bibitem{bell}
S.~Belluchi and V.~Faraoni, Nucl.\ Phys.\ B\ {\bf 640}, 453
(2002).

\bibitem{cal}
 A, Helmi and H.~Vucetich, Phus.\ Lett.\ A {\bf 230}, 153
(1997);\\
A.~E.~Motter and P.~S.~Letelier, Phys.\ Rev.\ D.\ {\bf 65}, 068502
(2002).

\bibitem{roch}
T.~M.~Rocha Filho {\it{et al.}}, Int.\ J.\ Theor.\ Phys.\ {\bf 39}
1933 (2000).

\bibitem{cast}
M.~A. Castagnino, H.~Giacomini and L.~Lara, Phys.\ Rev.\ D {\bf
61}, 107302 (2000).

\bibitem{Hilbert} R. Courant and D. Hilbert, {\it {Methods of
Mathematical Physics}}, Vol. 2, Cambridge University Press.

\bibitem{Chaos}
G.~K.~Savvidy, Phys.\ Lett.\ B {\bf 130}, 303 (1983);\\
W.~H.~Steeb, J.~A.~Louw and C.~M.~Villet, Phys.\ Rev.\ D\ {\bf
33}, 1174 (1986);\\
P.~Dahlqvist and G.~Russberg,  Phys.\ Rev.\ Lett.\  {\bf 65}, 2837
(1990).

\bibitem{cyclic}
P.~J.~Steinhardt and N.~Turok, Science {\bf 296}, 1436
(2002);\\
L.~Baum and P.~H.~Frampton, Phys. Rev. Lett. {\bf 98}, 071301
(2007) [arXiv:hep-th/0610213];\\
E.~N.~Saridakis, [arXiv:hep-th/0710.5269].

\bibitem{Chaos.cosmol}
E.~Calzetta and C.~E.~Hasi, Class.\ Quant.\ Grav.\ {\bf 10}, 1825
(1993) [arXiv:gr-qc/9211027];\\
L.~Bombelli, F.~Lombardo and M.~Castagnino, J. Math. Phys.\ {\bf
39}, 6040 (1998) [arXiv:gr-qc/9707051].

\bibitem{szy}
M.~Szydlowski, W.~Czaja and A.~Krawiec, [arXiv:astro-ph/0401293].



 \end{thebibliography}
\end{document}